\documentclass[graybox]{svmult}

\newcommand{\beq}{\begin{equation}}
\newcommand{\eeq}{\end{equation}}
\newcommand{\bea}{\begin{eqnarray}}
\newcommand{\eea}{\end{eqnarray}}

\newcommand{\Mpl}{M_{\rm P}}
\newcommand{\Madm}{M_{\rm ADM}}
\newcommand{\ellp}{R_{\rm P}}

\newcommand{\tpl}{t_{\rm P}}

\usepackage{mathptmx}       
\usepackage{helvet}         
\usepackage{courier}        
\usepackage{type1cm}        
\usepackage{makeidx}         
\usepackage{graphicx}        
\usepackage{multicol}        
\usepackage[bottom]{footmisc}

\makeindex             

\begin{document}

\title*{The Generalized Uncertainty Principle and Higher Dimensions: Linking
Black Holes and Elementary Particles}
 \titlerunning{The Generalized Uncertainty Principle and Higher Dimensions} 
\author{B. J. Carr}
\institute{
School of Physics and Astronomy, Queen Mary University  of London, Mile End Road, London E1 4NS, UK, \email{B.J.Carr@qmul.ac.uk}}
%
%
\maketitle

\abstract{
Black holes play an
 important role in linking microphysics with macrophysics, with those of the Planck mass ($\Mpl \sim10^{-5}$g) featuring in any theory of quantum gravity.  In particular, the Compton-Schwarzschild correspondence posits a smooth transition between the Compton wavelength ($R_{\rm C} \propto 1/M$) below the Planck mass and the Schwarzschild radius ($R_{\rm S} \propto M$) above it.
The duality between $R_{\rm C}$ and $R_{\rm S}$ implies a form of the  Generalized Uncertainty Principle (GUP) and suggests 
that elementary particles may be sub-Planckian black holes. The simplest possibility
is that the ADM mass has the form $M + \beta \Mpl^2/M$ for some constant $\beta$ and this model can be extended to charged and rotating black holes, clearly relevant to elementary particles.
Another possibility
is that sub-Planckian black holes may
arise in loop quantum gravity and this explicitly links black holes and elementary particles.  Higher dimensions may modify both proposals.
If there are $n$ extra dimensions,  all with the same compactification scale,  one expects $R_{\rm S} \propto M^{1/(1+n)}$ below this scale
but $R_{\rm C}$ depends on the form of the higher-dimensional wave-function.
If it is spherically symmetric,
then $R_{\rm C} \propto M^{-1}$, so
duality is broken
 and the Planck mass is reduced,
allowing the possibility of TeV quantum gravity. 
If the wave-function is pancaked in the extra dimensions, 
$R_{\rm C} \propto M^{-1/(1+n)}$ and so duality 
is preserved but 
the Planck mass is unchanged.
}

\section{Introduction}
\label{sec:1}

Whatever final theory amalgamates
 relativity theory and quantum mechanics,  it is likely to involve two features: (1) 
what is termed the 
Black Hole Uncertainty Principle (BHUP)
 correspondence; and (2) the existence of extra dimensions 
on sufficiently small scales.  Both features are expected to become important at the Planck length, $R_{\rm P} \sim 10^{-33}$cm, and possibly on much larger scales.
It is striking that black holes are involved in both these features and indeed there are many other ways in which these objects provide a link between macrophysics and microphysics \cite{carr 2018}.

As regards feature (1), the 
 duality under the transformation $M \rightarrow M_{\rm P}^2/M$ between the Compton wavelength for a particle of mass $M$, $R_{\rm C} = \hbar/(Mc)$, 
and the Schwarzschild radius for a black hole of mass $M$, $R_{\rm S} = 2GM/c^2$,
suggests a unified Compton-Schwarzschild expression with a smooth minimum in the ($M,R$) plane.
This implies that elementary particles may in some sense be sub-Planckian black holes.
This proposal goes back to the 1970s, when it was motivated in the context of strong gravity theories.

As regards feature (2), 
if there are $n$ extra spatial dimensions compactified on some scale $R_E$, then $R_{\rm S}$ scales as $R^{1/(1+n)}$ for $R < R_E$, 
leading to the possibility of TeV quantum gravity and 
black hole production at accelerators if $R_{\rm C}$ scales as $M^{-1}$ for $R < R_E$.
However,  the
higher-dimensional Compton wavelength depends on the form of the $(3+n)$-dimensional wavefunction
and in some circumstances one might expect
$R_{\rm C} \propto M^{-1/(1+n)}$ for $R < R_E$. This preserves the duality between $R_{\rm C}$ and $R_{\rm S}$ but
TeV quantum gravity is precluded.
Nevertheless, the extra dimensions could still have consequences for the detectability of black hole evaporations and the enhancement of pair-production at accelerators on scales below $R_E$. 

The plan of this paper is as follows. Sec.~2 discusses the 
BHUP correspondence 
in general terms.  Sec.~3 applies the correspondence to black holes in Loop Quantum Gravity (LQG), 
this being the first historical study of this kind.  Sec.~4 then considers the simplest application of the BHUP correspondence: the`$M+1/M$' Schwarzschild model and its extension to charged and rotating black holes.
 Higher-dimensional black holes are discussed in Sec.~5 and
some concluding remarks about the connection between particles and black holes are made in Sec.~6.  

\section{The Black Hole Uncertainty Principle Correspondence}
\label{sec:2}

\begin{figure}[b]
\sidecaption
\includegraphics[scale=.22]{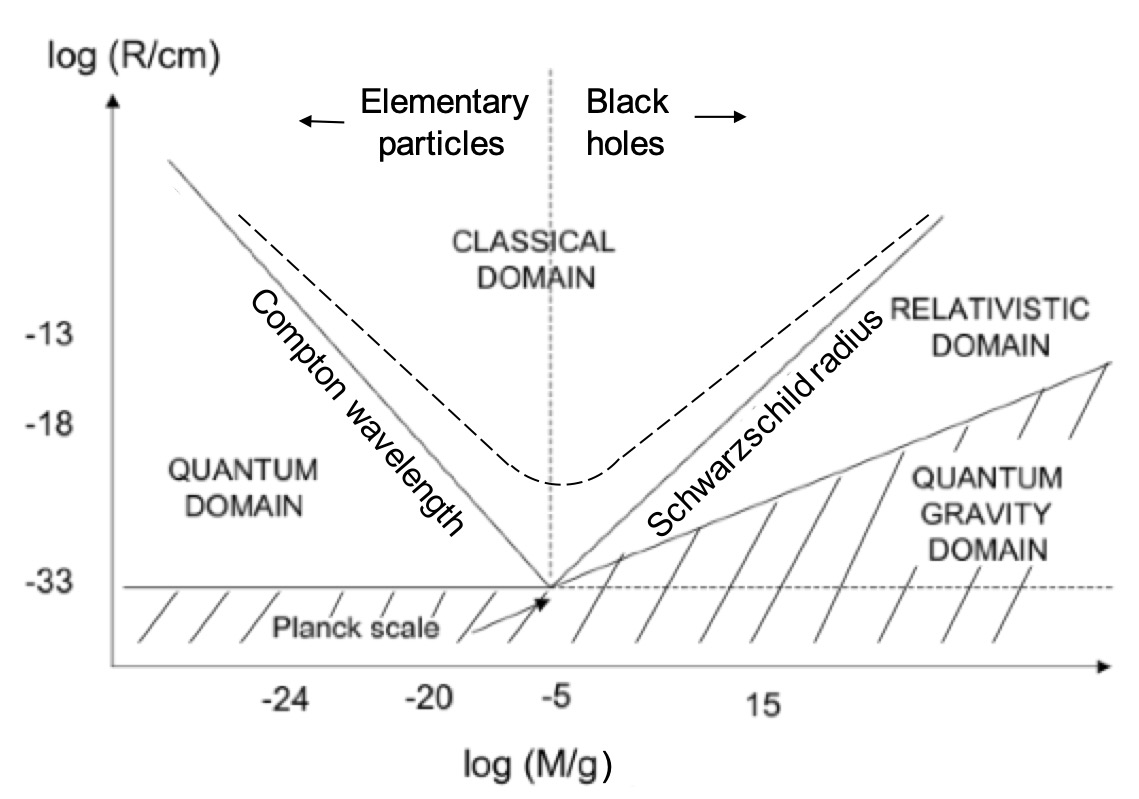}
\caption{
Division of ($M,R$) diagram into classical, quantum, relativistic and quantum gravity domains. The boundaries are specified by the Planck density, Compton wavelength and Schwarzschild radius. From Ref.~\cite{carr 2018}.} 
\label{MR}
\end{figure}

A key feature of the microscopic domain is the (reduced) Compton wavelength for a particle of rest mass $M$, 
with the region 
$R<R_{\rm C}$ in the $(M,R)$ diagram of Fig.~\ref{MR} being regarded as the ``quantum domain''.
A key feature of the macroscopic domain is the Schwarzschild radius for a body of mass $M$, 
with the region $R<R_{\rm S}$ being regarded as the ``relativistic domain''.
The Compton and Schwarzschild lines intersect at around the Planck scales,
\begin{equation}
R_{\rm P} = \sqrt{ \hbar G/c^3} \sim 10^{-33} \mathrm {cm}, \quad
M_{\rm P} = \sqrt{ \hbar c/G} \sim 10^{-5} \mathrm g \, ,
\end{equation}
and divide the $(M,R)$ diagram
 into three regimes, which we label quantum, relativistic and classical. 
There are several other interesting lines in the figure. The vertical line $M=M_{\rm P}$ marks the division between elementary particles ($M <M_{\rm P}$) and black holes ($M > M_{\rm P}$), since the size of a black hole is usually required to be larger than the Compton wavelength associated with its mass. The horizontal line $R=R_{\rm P}$ is significant because quantum fluctuations in the metric should become important below this \cite{wheeler}.
Quantum gravity effects should also be important whenever the density exceeds the Planck value, $\rho_{\rm P} = c^5/(G^2  \hbar) \sim 10^{94} \mathrm {g \, cm^{-3}}$, corresponding to the sorts of curvature singularities associated with the big bang or the centres of black holes.
 This implies $R < R_{\rm P}(M/M_{\rm P})^{1/3}$, which is well above the $R = R_{\rm P}$ line 
for $M \gg M_{\rm P}$, so the shaded region specifies the `quantum gravity' domain. 

The Compton and Schwarzschild lines transform into one another under the 
transformation $M \rightarrow M_{\rm P}^2/M$,
which suggests some connection between elementary particles and black holes. 
This relates to what is termed 'T-duality' in string theory
and maps momentum-carrying states to winding states \cite{Zwiebach:2004tj}. 
Although the Compton and Schwarzschild boundaries correspond to straight lines in the logarithmic plot of Fig.~\ref{MR}, this form presumably breaks down near the Planck point due to quantum gravity effects. One might envisage two possibilities: either there is some form of 
critical point at the Planck scale, so that the separation between particles and black holes is maintained \cite{mijmpn}, or there is a smooth minimum, as indicated by the broken line in Fig.~\ref{MR}, so that the Compton and Schwarzschild lines 
merge \cite{CaMuNic:2014}. Which alternative applies has important implications for the relationship between elementary particles and black holes.
 
One way of smoothing the transition between the Compton and Schwarzschild lines is to invoke
some form of 
unified expression which asymptotes to  the Compton wavelength and Schwarzschild radius in the appropriate regimes \cite{Ca:2014b}. 
The simplest such expression would be 
\begin{equation} \label{BHUP}
R_{\rm CS} =  \frac{\beta \hbar}{Mc} + \frac{2GM}{c^2}  \, ,
\end{equation}
where $\beta$ is a dimensionless constant.
In the super-Planckian regime, this becomes
\begin{equation} \label{GEH1} 
R_{\rm S}' = \frac{2GM}{c^2} \left[ 1 + \frac{\beta}{2} \left(\frac{ M_{\rm P}}{ M} \right)^2 \right] \quad (M \gg M_{\rm P}) \, ,
\end{equation}
with the second term corresponding to a small correction to the usual Schwarzschild expression.  In the sub-Planckian regime, it becomes
\begin{equation} \label{GUP1}
R_{\rm C}'  = \frac{\beta \hbar}{Mc} \left[1  + \frac{2}{\beta}  \left( \frac{M}{M_{\rm P}} \right)^2 \right]  \quad (M \ll M_{\rm P})  \, ,
 \end{equation}
with the second term corresponding to a small correction to the usual expression for the Compton wavelength.
More generally, 
one might consider any unified expression $R_{\rm C}'(M) \equiv R_{\rm S}'(M)$ which has the asymptotic behaviour $\beta \hbar/(Mc)$ for $M \ll M_{\rm P}$ and $2GM/c^2$ for $M \gg M_{\rm P}$. 

An expression of the form (\ref{GEH1}) arises in 
the quantum N-portrait model of Dvali {\it et al.} \cite{dvali},
which regards a black hole as a weakly-coupled Bose-Einstein condensate of gravitons.  From holographic considerations, the number of gravitons in the black hole is $N  \approx M^2/ \Mpl^2$
and one can then argue that the black hole radius is~\cite{frassino}
\beq
R_{\rm CS} \approx 
\frac{2GM}{c^2} \left(1 +\frac{\beta}{2N}\right) 
 \quad (M > \Mpl) \, ,
\label{case2}
\eeq
which is equivalent to Eq.~(\ref{GEH1}).
An expression of the form (\ref{GUP1}) also arises in the context of the Generalized Uncertainty Principle (GUP). 
This is because it can be argued that the Uncertainty Principle should be modified to the form~\cite{adler} 
 \begin{equation}
\Delta x = \frac{\hbar}{\Delta p} + \alpha \, \frac{R_{\rm P}^2 \, \Delta p}{\hbar} \, ,
\label{GUP} 
\end{equation}
where $\alpha$ is a dimensionless constant. 
The first term represents the uncertainty in the position due to the momentum of the probing photon and leads to the usual 
expression for the Compton wavelength if one substitutes $\Delta x \rightarrow R$ and $\Delta p \rightarrow c M$. The second term 
 represents the gravitational effect  of the  probing photon
and  is much smaller than the first term for $\Delta p \ll cM_{\rm P}$. 
Variants of Eq.~(\ref{GUP}) 
are also motivated by string theory~\cite{veneziano}, non-commutative quantum mechanics~\cite{scardigli}, general minimum length considerations \cite{maggiore}, polymer corrections in the structure of spacetime in LQG~\cite{ashtekar} and some approaches to quantum decoherence~\cite{kay}.

The GUP is usually restricted to the sub-Planckian domain ($M < \Mpl$). However,  if we rewrite Eq.~(\ref{GUP}) using $\Delta x \rightarrow R$ and $\Delta p \rightarrow c M$ even in the super-Planckian regime, 
we obtain a revised Compton wavelength which applies for all $M$:
 \begin{equation}
R_{\rm CS} =   \frac{\hbar}{Mc} + \alpha \, \frac{GM}{c^2} \, .
\label{GUP2}
 \end{equation}
This resembles Eq.~(\ref{BHUP}) except that the constant is associated with the second term.  
This suggests that  there is a different kind of positional uncertainty for an object larger than the Planck mass, related to the size of a black hole. 
This is not unreasonable since the usual Compton wavelength is below the Planck length 
here and also an outside observer cannot localize an object on a scale smaller than its Schwarzschild radius. 
This is termed the Black Hole Uncertainty Principle (BHUP) correspondence \cite{Ca:2014b} 
or the Compton-Schwarzschild correspondence when discussing an interpretation in terms of extended de Broglie relations \cite{Lake:2015pma}.

Strictly speaking, Eqs. (\ref{BHUP}) and (\ref{GUP2}) are consistent only if $\alpha = 2$ and $\beta =1$ but that would leave no free parameter at all.  Therefore an interesting issue is whether one should associate the free constant in $R_{\rm CS}$ with the $1/M$ term, as in Eq.~(\ref{BHUP}), or the $M$ term,  as in Eq.~(\ref{GUP2}).  Here we adopt the former approach, on the grounds that the expression for the Schwarschild radius is exact, whereas there is some ambiguity in the meaning of the Compton scale.
However,  for comparison with the GUP literature , we still need to identify an effective value of $\alpha$ and a simple rescaling of the relationship between $\Delta x$ and $R$ suggests $\alpha= 2/\beta$.  Another approach is to identify $\Delta p$ with $1/M$ rather than $M$ for $M >M_{\rm P}$ and Eq.~(\ref{GUP}) then equates $\alpha$ and $\beta$ directly.  
One might even argue that 
$\Delta p$ has the form $(M + 1/M)^{-1}$, in which case 
$\Delta x \sim 1/M$ and $\Delta p \sim M$ in the particle case ($M < M_{\rm P}$) and $\Delta x \sim M$ and $\Delta p \sim 1/M$ in the black hole case ($M >M_{\rm P}$).  This would be consistent with  the extended de Broglie relations \cite{Lake:2015pma}. 
\if
{\bf One might also consider a relationship of the form}
 \begin{equation}
 \Delta x \, \Delta p \sim 
1  + (\delta p)^2 + (\delta x)^2 \, 
 \end{equation}
(omitting coefficients),
with the last term corresponding to what is termed the Extended Uncertainty Principle~\cite{mureika2019}. {\bf However, the effect of this term would only become apparent on large scales, so it irrelevant to the present considerations.}
\fi

\begin{figure}[b]
\sidecaption
\includegraphics[scale=.12]{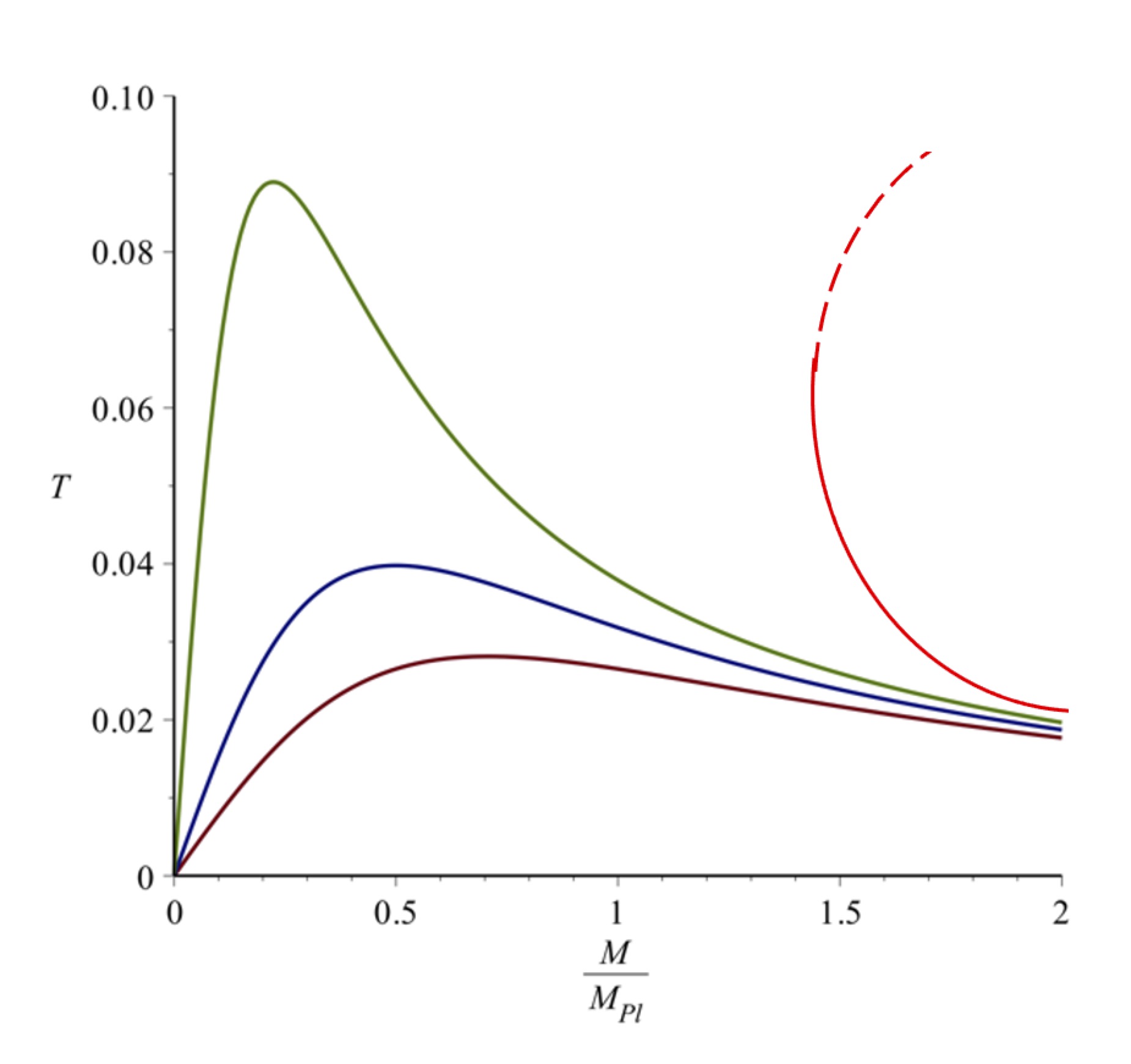}\caption{
 Hawking temperature (in Planck units) from Eq.~(\ref{GEH1}) and surface gravity argument as a function of $M/\Mpl$ for $\beta = 1$ (bottom), $\beta = 0.5$ (middle) and $\beta = 0.1$ (top).  Also shown on the right is the Adler  prediction.
From Ref.~\cite{CaMuNic:2014}.
\label{fig1}
}
\end{figure}

In the standard picture, one can  calculate the black hole temperature from the Uncertainty Principle by 
identifying it
with a multiple $\eta$ of  $\Delta p$. This gives \cite{hawking}
\begin{eqnarray}
kT = \eta c \Delta p = \frac{ \eta  \hbar c}{\Delta x} = \frac{\eta c^2  \Mpl^2}{ 2M} \, ,
\label{temper}
\end{eqnarray}
which is precisely the Hawking temperature 
if we take $\eta = 1/(4\pi)$. 
If one adopts the GUP but assumes  the usual 
black hole size,  one obtains
the Adler form~\cite{chen}
\begin{equation}
kT = {\eta M c^2  \over \alpha} \left(1\pm \sqrt{1- \frac{\alpha \Mpl^2}{M^2}} \right) \, .
\label{adlertemp2}
\end{equation}
The negative sign 
gives a small perturbation to the standard Hawking temperature
\begin{equation}
kT \approx
{\eta \Mpl^2 c^2 \over 2 M}  \left[ 1 - {\alpha \Mpl^2 \over 4M^2} \right] \quad (M \gg \Mpl) \, 
\label{adler3}
\end{equation}
at large $M$.  However, the solution becomes complex when $M$ falls below $\sqrt{\alpha} \, \Mpl$, corresponding to a minimum mass,
and it then connects to the positive branch of Eq.~(\ref{adlertemp2}). 
This form is indicated by the curve on the right of Fig.~\ref{fig1}.

Equation~(\ref{adlertemp2}) is inconsistent with the BHUP correspondence since this
also modifies the relationship between the black hole radius $\Delta x$ and $M$.
If we adopt Eq.~(\ref{GEH1}) instead, then the surface gravity argument gives a temperature 
\bea
kT  = \frac{\Mpl^2 c^2}{4\pi M(2 + \beta \Mpl^2/M^2)} \approx
\left\lbrace
\begin{array}{ll}
\frac{\Mpl^2 c^2}{8\pi M} 
\left[ 1- \frac{\beta}{2}  \left( \frac{\Mpl}{M} \right)^{2} \right] 
 & (M \gg \Mpl) \\
\frac{M c^2}{4\pi \beta} 
\left[ 1- \frac{2}{\beta} \left( \frac{M}{\Mpl} \right)^2 \right] 
& (M \ll \Mpl)  \, .
\end{array}
\right.
\label{hawktemp}
\eea
This  is plotted in Fig.~\ref{fig1} and is very different from the Adler form. 
As $M$ decreases,
the temperature reaches a maximum of around $T_{\rm P}$
and then goes to zero  as $M \rightarrow 0$. 

An important caveat is that
Eq.~(\ref{GUP}) assumes that the two uncertainties add linearly.
 On the other hand, 
since they are independent, it might be more natural to assume that they add quadratically \cite{CaMoPr:2011}:
 \begin{equation}
 \Delta x = \sqrt {(\hbar/ \Delta p)^2 + (\alpha R_{\rm P}^2 \Delta p/ \hbar)^2} \, .
 \label{quad2}
 \end{equation}
We refer to Eqs.~(\ref{GUP}) and (\ref{quad2}) as the {\it linear} and {\it quadratic} forms of the GEP, respectively. The latter corresponds to a 
unified expression
\begin{equation}
R_{\rm CS} = \sqrt{ (\beta \hbar /Mc)^2  + (2GM/c^2)^2 } \, , 
\label{quadEH}
\end{equation}
where we have we have again introduced $\beta$.  This leads to the approximations
 \begin{equation}
R _{\rm S}' \approx \frac{2GM}{c^2} \left[ 1  + \frac{\beta^2}{8} \left( \frac{M_{\rm P}}{M} \right)^4 \right] \quad (M \gg M_{\rm P})
\end{equation}
 and
 \begin{equation}
R _{\rm C}' \approx \frac{\beta \hbar}{Mc}  \left[ 1  + \frac{2}{\beta^2} \left( \frac{M}{M_{\rm P}} \right)^4 \right] \quad (M \ll M_{\rm P})
\label{quadC} \, .
\end{equation}
These might be compared to the {\it exact} expressions in the linear case, given by Eqs.~(\ref{GEH1}) and (\ref{GUP1}).
As we now show, a model inspired by LQG permits the existence of a black hole whose horizon size has precisely the form (\ref{quadEH}).

\section{Loop Black Holes }

Loop Quantum Gravity is based on a canonical quantization of the Einstein equations, written in terms of the Ashtekar variables~\cite{LQGgeneral}.
One feature of this is 
that area is quantized, with its smallest possible value being
\begin{equation}
A_{\mathrm{min}} = 4 \pi \sqrt{3} \, \gamma \, R_{\rm P}^2 \, ,
\end{equation}
where $\gamma$ is 
the Immirzi parameter and of order $1$. 
The quantity
$a_o \equiv A_{\rm min}/8\pi$,
together with 
the dimensionless polymeric
parameter  $\delta$,
determines the deviation from classical theory.

One version of LQG,  using the mini-superspace approximation, gives rise to
cosmological solutions which
resolve the initial singularity problem \cite{Bojowald}.
Another version gives the loop black hole (LBH) solution
 \cite{poly} and this 
replaces the singularity in the Schwarzschild solution 
with another asymptotically flat region. 
The metric
depends only on the combined dimensionless parameter $\epsilon \equiv \delta \gamma$, which must be small,
and can be expressed as
\begin{eqnarray}
ds^2 = - G(r) c^2 dt^2 + \frac{dr^2}{F(r)} + H(r) (d \theta^2 + \sin^2 \theta d \phi^2)
\label{g}
\end{eqnarray}
with 
\begin{eqnarray}
G(r) = \frac{(r-r_+)(r-r_-)(r+ r_{*})^2}{r^4 +a_o^2},  \quad
F(r) = \frac{(r-r_+)(r-r_-) r^4}{(r+ r_{*})^2 (r^4 +a_o^2)}\,.
\label{statgmunu}
\end{eqnarray}
Here $r_+ = 2GM/c^2$ and $r_-= 2G M P^2/c^2$ are the outer and inner horizons, respectively, and $r_* \equiv \sqrt{r_+ r_-} = 2GMP/c^2$, where $M$ is the black hole mass and 
\begin{equation}
P \equiv  \frac {\sqrt{1+\epsilon^2} -1}{\sqrt{1+\epsilon^2} +1}   \approx \epsilon^2 /4 \ll 1
\end{equation}
is called the polymeric function. 
In the limit $r \to \infty$ one has 
\begin{eqnarray}
G(r) \to 1-\frac{2 G {\cal M}}{c^2 r} (1 - \epsilon^2), \quad
F(r) \to 1-\frac{2 G {\cal M}}{c^2 r}, 
\end{eqnarray}
where 
${\cal M} \equiv M (1+P)^2$
is the ADM mass (i.e.  the mass measured as $r \rightarrow \infty$).
The function $H(r)$ in Eq.~(\ref{g})
 is not $r^2$ (as in the Schwarzschild case) but
\begin{eqnarray}
H(r) = r^2 + \frac{a_o^2}{r^2} \quad \Rightarrow \quad R \equiv \sqrt{r^2 + \frac{a_o^2}{r^2}} \, .
\label{rho}
\end{eqnarray}
Here $R$ is the {\it physical} radial coordinate,  in the sense that the circumference function $2 \pi R$. 
As $r$ decreases from infinity to zero, $R$ first decreases from infinity to a minimum value of $\sqrt{2a_0}$ at $r=\sqrt{a_0}$ and then increases again to infinity. 
In particular, the value of $R$ associated with the outer event horizon is
\begin{eqnarray}
R_{\rm S}' 
= \sqrt{ \left( \frac{2 G M}{c^2} \right)^2 + \left( \frac{a_o c^2}{2 G M} \right)^2 }
\,.
\label{loopEH}
\end{eqnarray}
This corresponds to Eq.~(\ref{quadEH}) if 
$\beta = a_o c^2/G$.
The important physical implication of Eq.~(\ref{rho}) is that central singularity of the Schwarzschild solution is replaced with another asymptotic region, so the collapsing matter bounces and the black hole becomes part of a wormhole.
The fact that a purely geometrical condition in LQG
 implies the quadratic version of the GUP suggests some deep connection between general relativity and quantum theory.  The duality between the two asympotic spaces also suggests a link between elementary particles with $M \ll \Mpl$ and black holes with $M \gg \Mpl$ \cite{modesto}, which is clearly relevant to the theme of this paper.

The temperature implied by the black hole's surface gravity is 
\begin{eqnarray}
T \propto \frac{GM}{R_{\rm S}'^2} \propto 
\left\lbrace
\begin{array}{ll}
M^{-1} & (M \gg M_{\rm P}) \\
M^3 & (M \ll M_{\rm P}) \, .
\label{GUPtemp}
\end{array}
\right.
\end{eqnarray}
However,  if one calculates the temperature using the GUP expression for $\Delta p$, 
one obtains
\begin{equation}
k T \approx
\left\lbrace
\begin{array}{ll}
 {\eta \hbar c^3 \over 2 G M} \left[ 1 - \frac {\beta^2}{8} \left( {M_{\rm P} \over M} \right)^{4} \right]  & (M \gg M_{\rm P}) \\
{\eta M c^2 \over  \beta} \left[ 1 - \frac{2}{ \beta^2} \left( {M \over M_{\rm P}} \right)^{4} \right]   & (M \ll M_{\rm P}) \, . 
\end{array}
\right.
\label{subtemp2}
\end{equation}
This is similar to Eq.~(\ref{hawktemp})
but inconsistent with Eq.~(\ref{GUPtemp}) in the sub-Planckian regime.
The source of the discrepancy 
is that there are two asymptotic spaces -- one on each side of the wormhole throat -- and the temperature is different in these.  Observer only detect radiation from the horizon on their side of the throat, so 
the inner horizon with respect to $r = \infty$ corresponds to the outer horizon with respect to $r=0$ \cite{CaMoPr:2011}. 
For $M < \sqrt{\beta/2} \, M_{\rm P}$, 
$T \propto M^{3}$ in our space and $T \propto M $ in the other space,
which explains the predictions of Eqs.~(\ref{GUPtemp}) and (\ref{subtemp2}).  For $\sqrt{\beta/2} \, M_{\rm P} < M < P^{-2} \sqrt{\beta/2} \, M_{\rm P}$,  
 $T \propto M^{-1}$ in our space and $T \propto M$ in the other space. 
For $M > P^{-2} \sqrt{\beta/2} \, M_{\rm P}$,  
$T \propto M^{-1}$ in our space and $T \propto M^{-3}$ in the other space. 

\section{Carr-Mureika-Nicolini approach }

This section describes a particular interpretation of the linear version of the BHUP correspondence, 
described  in my work with Mureika and Nicolini~\cite{CaMuNic:2014},
in which the Arnowitt-Deser-Misner (ADM) mass
is taken to be 
\begin{equation}
\Madm = M \left(1+\frac{\beta}{2} \frac{\Mpl^2}{M^2}\right) \, .
\label{newmass}
\end{equation}
This is equivalent to Eq.~(\ref{GEH1}) and
we noted a possible connection with the energy-dependent metric in the ``gravity's rainbow'' proposal \cite{rainbow} and with the QFT renormalization of mass
in the presence of stochastic metric fluctuations \cite{camacho}.
Putting $\hbar = c =1$, the Schwarzschild radius for the modified metric is 
\bea
R_{\rm S}' = \frac{2 \Madm} {\Mpl^2} 
 \approx
\left\lbrace
\begin{array}{ll}
2M/  \Mpl^2 & (M \gg \Mpl ) \\
\beta/M &(M \ll \Mpl ) \, 
\end{array}
\right.
\label{horizon}
\eea
 and the temperature is 
$kT  = \Mpl^2/(8\pi \Madm)$,
corresponding to Eq.~(\ref{hawktemp}).
This is not the only black hole metric allowed by the GUP.  This is illustrated by the discussion of LQG in Sec.~3 but there could be other relevant solutions in GR itself. However,  Eq.~(\ref{newmass}) gives the simplest such solution.  
\if
A possible explanation for the $M \ll M_{Pl}$ behaviour is that a decaying black hole makes a temporary transition to a (1+1)-D dilaton black hole when approaching the Planck scale,
since  this naturally encodes a $1/M$ term in its gravitational radius.
For according to t'Hooft \cite{thooft}, gravity might experience a (1+1)-D phase at the Planck scale due to spontaneous dimensional reduction, such a conjecture being further supported by studies of the fractal properties of a quantum spacetime at the Planck scale. 
An interesting application of the (1+1)-D approach to PBH formation was studied in collaboration with Athanasios Tzikas~\cite{tzikas}.
\fi

The black hole luminosity 
in this model is
$L = \xi^{-1} \Madm^{-2}$
where $\xi \sim t_\mathrm{P}/\Mpl^3$, so
the mass loss rate decreases when $M$ falls below $M_{\rm P}$ and the black hole never evaporates completely. There are two values of $M$ for which  the evaporation time ($\tau \sim M/L$) is comparable to the age of the Universe ($t_0 \sim 10^{17}$s).
One is super-Planckian,
$M_* \sim (t_0/ \xi)^{1/3} \sim 10^{15} {\rm g}$,  
this being the standard expression for the mass of a PBH evaporating at the present epoch.
The other is sub-Planckian, 
$M_{**} \sim  \beta^2 (\tpl/t_0) \Mpl  \sim 10^{-65}\beta^2 \, {\rm g}$,
although the mass cannot actually reach this value at the present epoch because
 the black hole is cooler than the CMB temperature
for $M < M_{\rm CMB} \sim 10^{-36} \beta$~g.
This leads to {\it effectively} stable relics of this mass. 

It is interesting to consider observational constraints on the parameter $\beta$ and these are discussed in Ref.~\cite{CMN2022}.  Within the GUP context, these only arise  in the microscopic domain and a variety of mechanical oscillator experiments imply $\alpha < 4 \times 10^4$ \cite{pik}. A similar bound arises from the AURIGA gravitational bar detector \cite{aur}.  Since  $\beta = 2/\alpha$, both bounds corresponding to a {\it lower} limit $\beta > 10^{-4}$.  Within the context of the BHUP correspondence, there are also constraints in the macroscopic domain from measuring the gravitational force between 100 mg masses with mm separation \cite{westphal} and these imply $\beta < 10^6$. 
Clearly these limits still allow a wide range of values for $\beta$.
One might also constrain $\beta$ by observations on astrophysical scales but in this domain the effects of the Extended Uncertainty Principle,  in which $\Delta x \, \Delta p \sim 1 +  (\delta x)^2$ rather than $1  +  (\delta p)^2$, becomes more relevant \cite{mureika2019}.

\begin{figure}[b]
\sidecaption
\includegraphics[scale=.21]{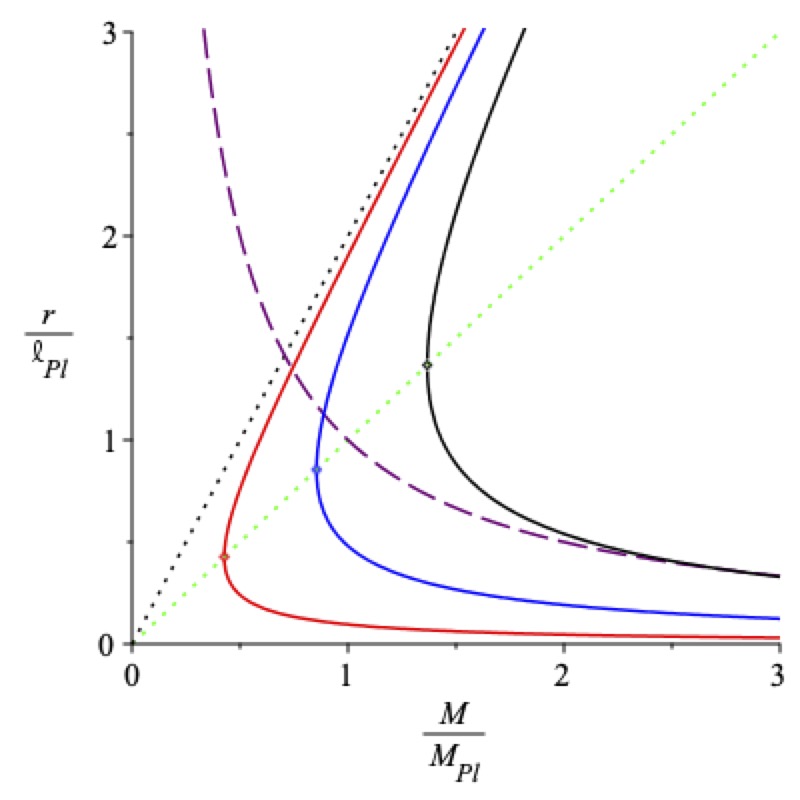}
\caption{
The solid curves show the  outer and inner horizons for a standard RN black hole with 
$n=5, 11, 16$ (left to right).  For each $n$,  the horizons meet at the extremal mass on the line $r=M/\Mpl^2$ (green dotted) and are bounded from above by the Schwarzschild radius $r_{\rm S} = 2M/\Mpl^2$ (black dotted line).  The Compton curve is shown by the dashed line and the inner horizon asymptotes to this for $n=16$.
Solutions with $11 < n < 16$ penetrate the sub-Planckian RN regime. From Ref.~\cite{CMMN}.}
\label{fig12}
\end{figure}

Recently we have extended this  work, together with Heather Mentzer, to charged and rotating black holes~\cite{CMMN}, since this is clearly relevant to elementary particles.
The standard Reissner-Nordstr\"om (RN) 
already exhibits features of the GUP-modified Schwarzschild solution. 
This is because the RN metric has
an outer (+) and inner (-) horizon at
\beq
r_\pm = \frac{M}{\Mpl^2}\left( 1\pm \sqrt{1-\frac{\alpha_e \Mpl^2 n^2}{M^2}}\right)
\approx 
\left\lbrace
\begin{array}{ll}
\frac{2M}{\Mpl^2} \left(1 - \frac{\gamma \Mpl^2}{M^2} \right) & (+) \\ 
\frac{2\gamma}{M} \left(1 + \frac{\gamma \Mpl^2}{M^2} \right) & (-)
\end{array}
\right.
\label{RNhorizon}
\eeq
where $ne$ is the black hole charge, $\gamma \equiv \alpha_e n^2/4$ with $\alpha_e \approx 1/137$ being the electric fine structure constant, and the last expression applies for a black hole which is far from extremal ($ M \gg \sqrt{\alpha_e} \, n \Mpl$).
The form of the outer and inner 
 horizons for different values of $n$ are shown  by the upper and lower  parts of the solid curves in Fig.~\ref{fig12}, respectively.  There are two asymptotic behaviors:
 the outer horizon correponds to 
Eq.~(\ref{BHUP}) but with a negative value of $\beta$; the inner horizon 
resembles the Compton expression and it asymptotes to the Compton wavelength itself for $n = 16$, this being the integer part of $\sqrt{2/\alpha_e}$.

For each $n$, the two horizons merge on the line  $r=GM$ (lower dotted curve) at the minimum value of $M$.
This corresponds to a sequence of ``extremal'' solutions (shown by the dots in Fig.~\ref{fig12}) with  a spectrum of masses 
$\sqrt{\alpha_e} \, n \Mpl$.
For given $n$, there are no solutions with  $M$ less than this since these would correspond to naked singularities. In particular, $n$ could be at most the integer part of $1/ \sqrt{\alpha_e}$ (i.e. $11 $) for a Planck-mass black hole.
As in the GUP case,  the temperature of the 
RN solution reaches a maximum and then goes to zero as $M$ tends to the limiting vaue $\sqrt{\alpha_e} \, n \Mpl$.  
One might want to associate elementary particles only with extremal solutions (since they are stable) but these states all have masses in the range $(0.1 - 1) \Mpl$, which is too large.  Also even extremal black holes may discharge through the Schwinger mechanism \cite{schwinger}.

The (standard) Compton line intersects the outer black hole horizon,
as required if one wants a smooth connection between particles and black holes, 
at the mass
\beq
M = \frac{\Mpl}{\sqrt{2-\alpha_e n^2}} \approx  \frac{\Mpl}{\sqrt{2- n^2/137}} \, .
\label{extreme}
\eeq
(This is also termed the self-completeness condition~\cite{mijmpn}.) 
For $n=0$, the intersect is
$\Mpl /\sqrt{2}$ but it increases with $n$ and tends to $\Mpl$ as $n \rightarrow \sqrt{137}$ (middle curve). This implies a constraint $n \leq 11$ on the charge of a self-complete RN black hole. The Compton line still intersects the {\it inner} horizon for $\sqrt{137} <  n < \sqrt{274} $, 
but these solutions  
penetrate the $r < \ellp$ region.

\begin{figure}[h]
\includegraphics[scale=0.11]{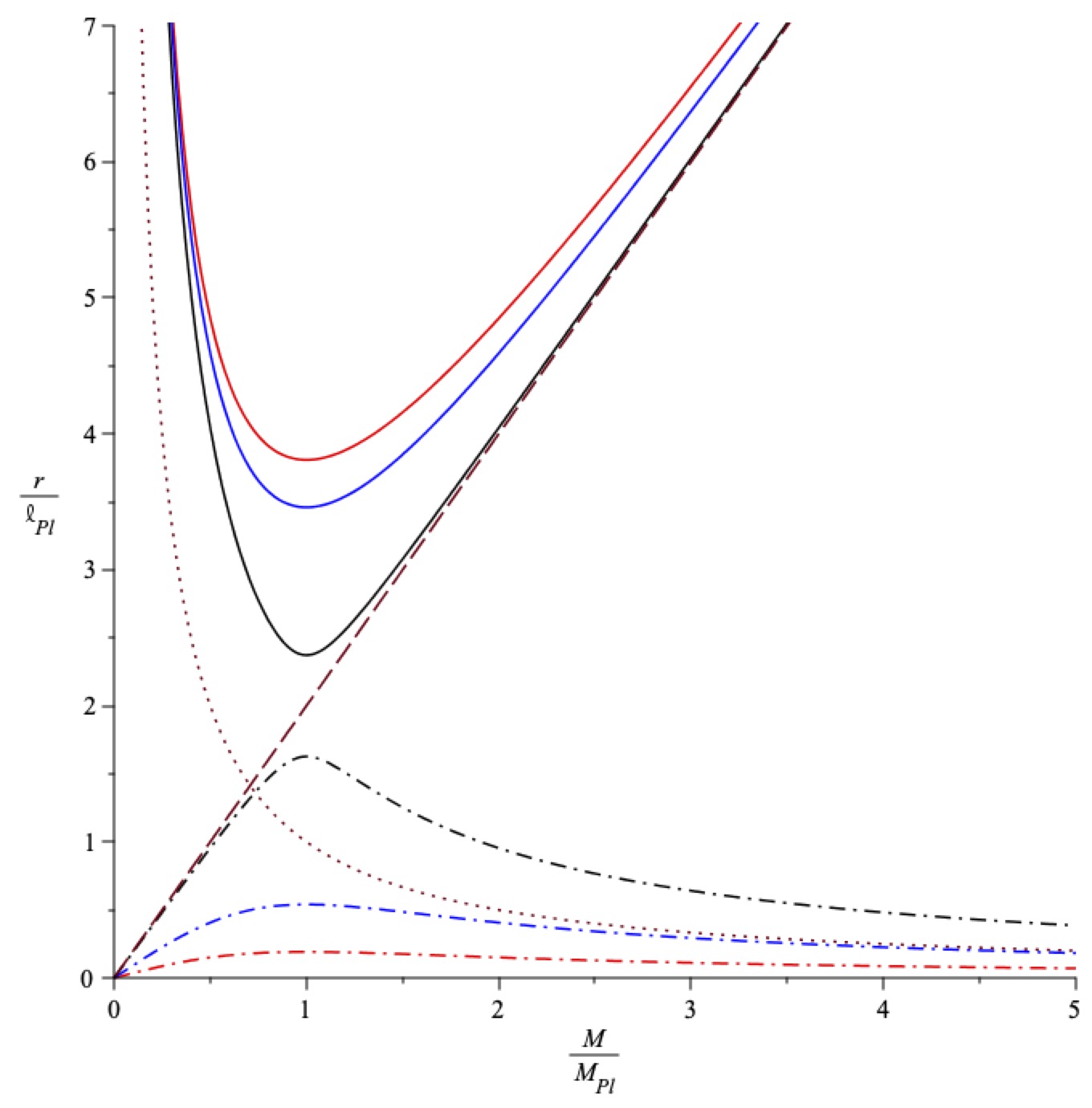}
\includegraphics[scale=0.10]{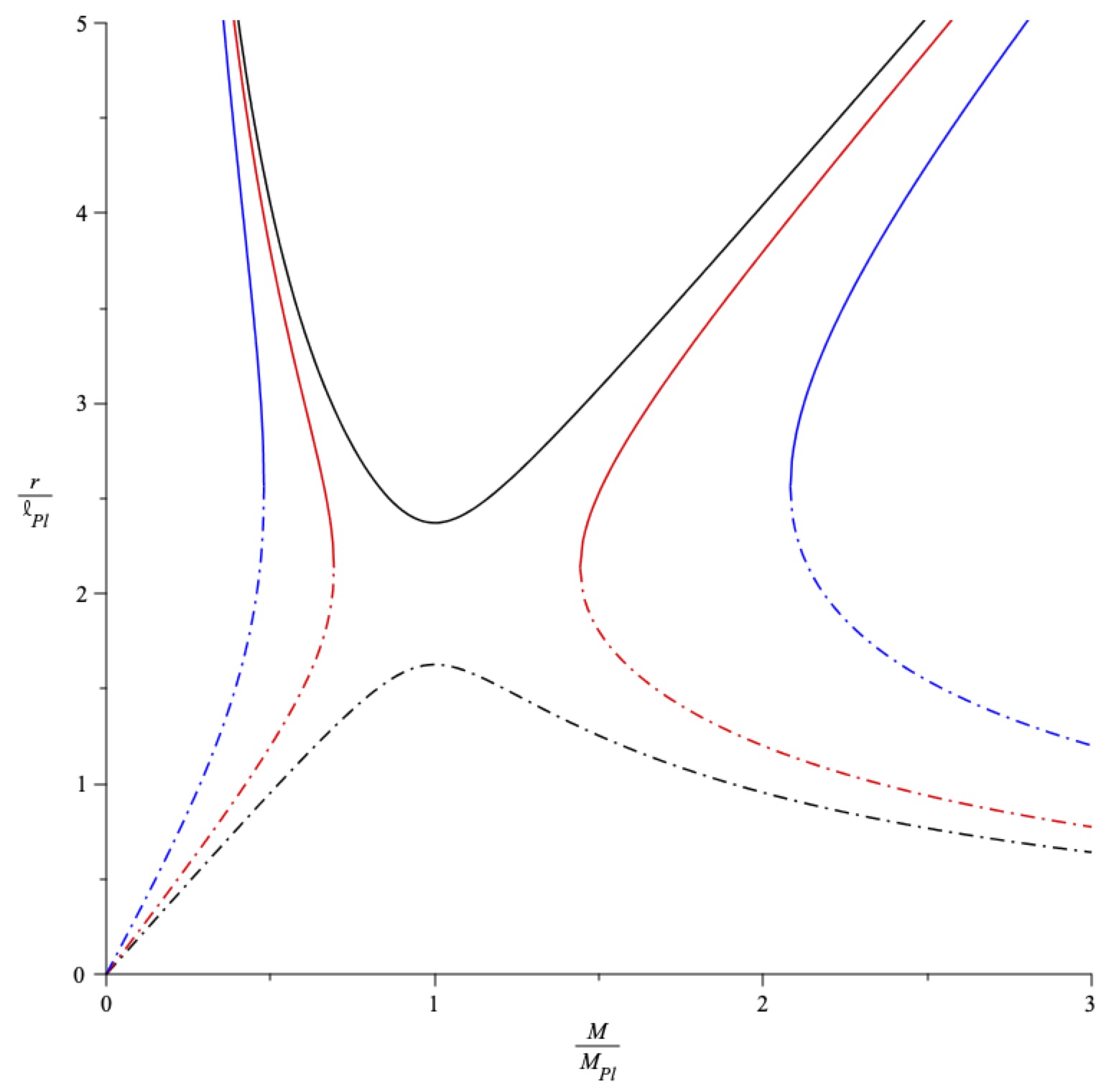}
\caption{Outer (solid) and inner (dash-dot) horizon size for GUP-RN black hole with $\beta =2$.
Left: 
Outer (top) and inner (bottom) horizons for 
$n=10$ (red), $n=16$ (blue) and $n=23$ (black).  The dashed/dotted lines show the usual Schwarzschild/Compton scales.
 The inner horizon is nearly asymptotic to the Compton wavelength at large $M$ for $n =16$.
There is a discontinuity when $n$ reaches 
$23$, this being close to an extremal solution.
Right: Outer (top) and inner (bottom) horizons for 
$n=23 $ (black), $n=25$ (blue) and $n=30$ (black).  
The horizons in this case have a maximum value of $M$ on the left and a minimum value on the right.
There are no black holes between these values. From Ref.~\cite{CMMN}.}
\label{PT_RNH}
\end{figure}

One can extend this model to the GUP-modified RN solutions by replacing $M$ with $M_{\rm ADM}$ given by Eq.~(\ref{newmass}). 
Providing 
$n <  \sqrt{2 \beta/ \alpha_e}$ for fixed $\beta$, the outer horizon behaves as in 
 the GUP-Schwarzschild case, 
with a continuous transition between the gravitational ($r_{\rm CS} \propto M$) and Compton ($r_{\rm CS}\propto M^{-1}$) scaling.  Also, $r_+$  has a minimum and $r_-$  has a maximum at
\beq
M = M_{\rm crit} \equiv \sqrt{\beta/2} \, \Mpl  \, , \quad  r_{\pm} = [\sqrt{2 \beta} \, {\pm} \, \sqrt{ 2 \beta - n^2 \alpha_e} \, ]  \,  \ellp \, .
\label{minhor}
\eeq
This is indicated by the curves on the left of Fig.~\ref{PT_RNH}.  In principle, the particle-like black holes can have arbitrarily low mass in this case.  However,  it is unclear that these solutions are candidates for {\it stable} particles since none of them are extremal, this possibility arising only in the limit $n_{max} =   \sqrt{2 \beta/ \alpha_e}$. 
For larger values of $n$,
the form of the solutions changes, as indicated by the curves on the right of Fig.~\ref{PT_RNH}. 
These represent super-Planckian black holes on the right (similar to the standard RN case with an extremal solution at the smallest value of $M$) and sub-Planckian particles on the left, with a mass gap in between.
Equivalently, for a given value of $n$, there is a critical value of $\beta = n^2 \alpha_e/2$ 
below which the solutions bifurcate 
and become separated by a mass gap.

The Kerr metric exhibits similar behaviour
but there is a critical spin ($n \hbar$)  
rather than a critical charge.
The extremal case corresponds to the spectrum of masses
$\sqrt{n} \, \Mpl$,
while the self-completeness condition corresponds to
\bea
M  =  
 \Mpl\sqrt{\frac{1+n^2}{2}} \, .
\label{sck}
\eea
This allows all values of $n$, whereas 
$n$ could not exceed  $[1/\sqrt{\alpha_e}] =11$ in the RN case.
In the GUP Kerr case,  an expression similar to Eq.~(\ref{minhor}) still applies
and there is a change in the form of the solutions for $n > 2\beta$.

\section{Higher-dimensional black holes}
\label{HDBH}

The black hole boundary in Fig.~\ref{MR} assumes there are three spatial dimensions but many theories suggest that the dimensionality could increase on small scales. 
Although the extra dimensions are often assumed to be compactified on the Planck length, there are also models
in which they are much larger. For example,  
the model of Arkani-Hamed et al. \cite{arkani} has $n$ extra spatial dimensions, all compactified on the same scale $R_E$.
If we assume that the standard expression for the Compton wavelength ($R_{\rm C} \propto M^{-1}$) still applies, 
then the masses with Compton and Schwarzschild scales  $R_E$ are
\begin{eqnarray}  \label{M_E} 
M_{\rm E} \equiv \frac{\hbar}{c R_E} \simeq  M_{\rm P}\frac{R_{\rm P}}{R_E} \, , \quad
M_E' \equiv \frac{c^2 R_E}{G} \simeq M_{\rm P} \frac{R_E}{R_{\rm P}} \,  .
\label{ME}
\end{eqnarray} 
For $R < R_E$,  the gravitational potential generated by a mass $M$ is
\begin{equation} \label{V1}
V_{\mathrm{grav}} =  \frac{G_D M}{R^{1+n}}  \, ,
\end{equation} 
where $G_D$ is the higher-dimensional gravitational constant and $D = 4+n$ is the total number of spacetime dimensions.  For $R>R_E$, 
one recovers the usual form, $V_{\mathrm{grav}} =  GM/R$ with $G = G_D/R_E^n$.
Thus the effective gravitational constants at large and small scales are different.

\begin{figure}[b]
\sidecaption
\includegraphics[scale=.14]{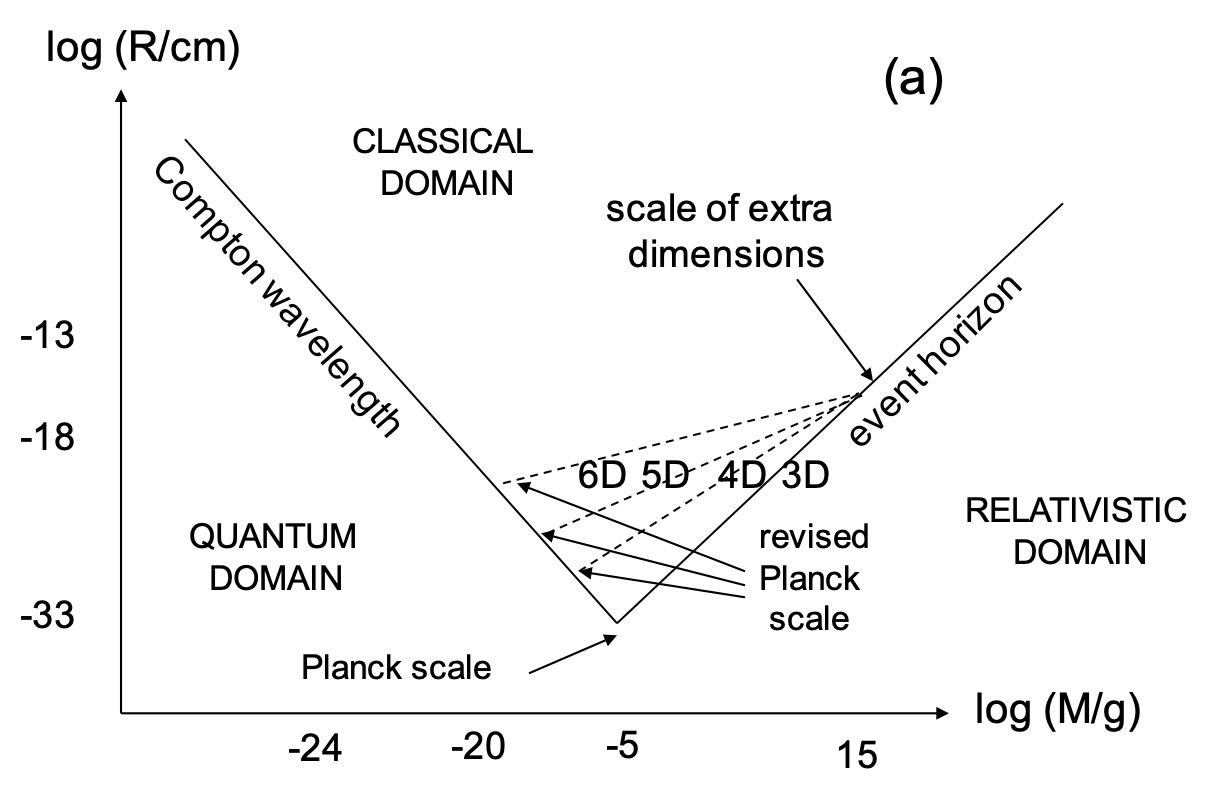}
\includegraphics[scale=.14]{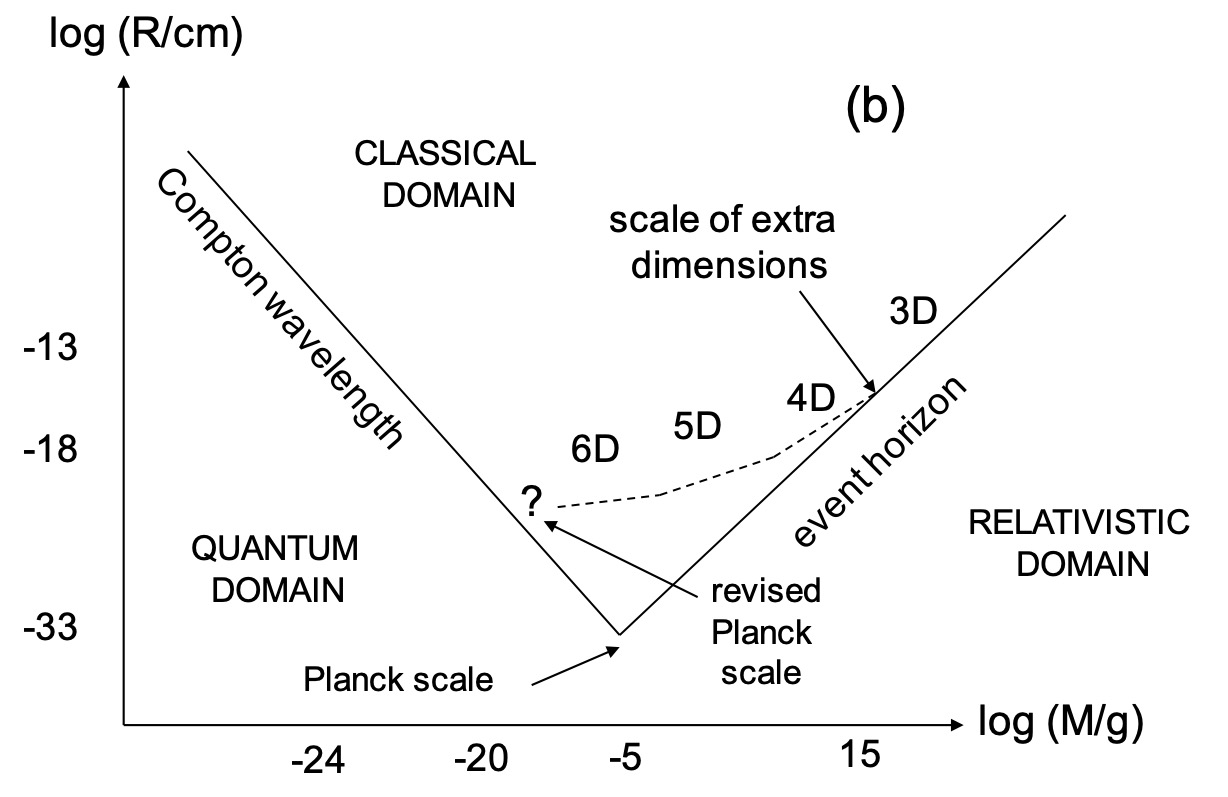}
\includegraphics[scale=.07]{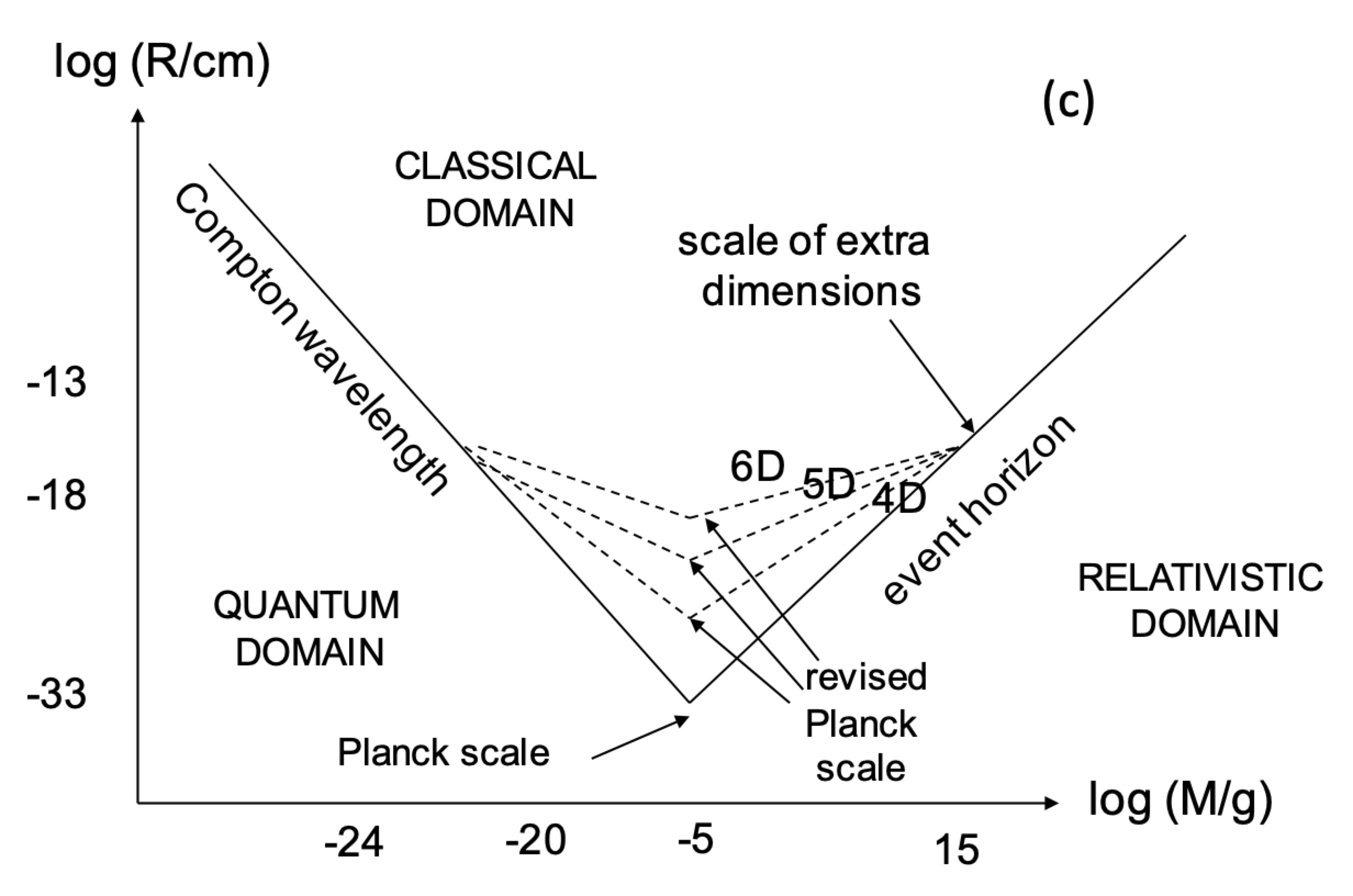}
\includegraphics[scale=.07]{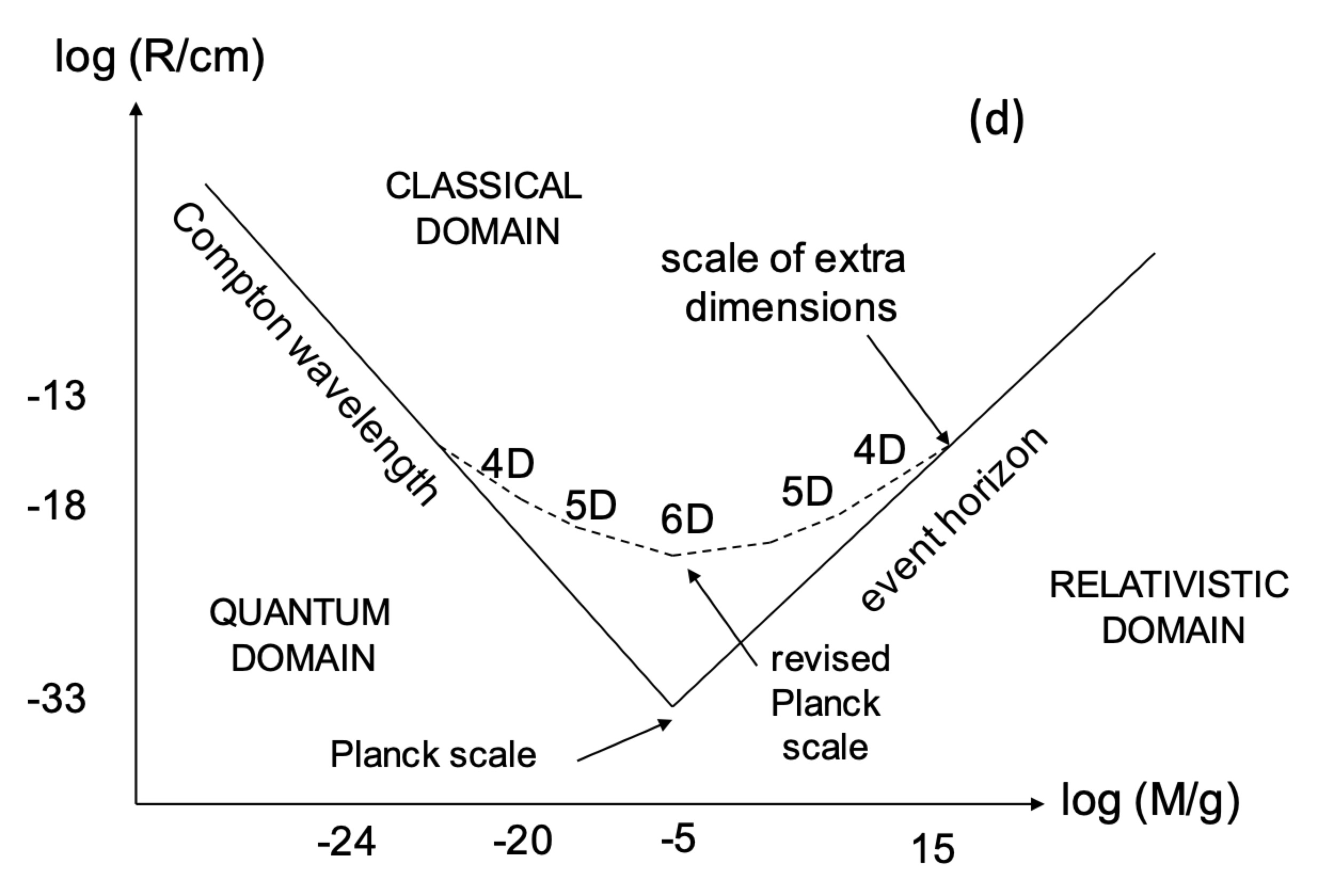}
\caption{
Modification of the Schwarzschild line  and Planck scales in the $(M,R)$ diagram for extra compact dimensions associated with (a) a single length scale or (b) a hierarchy of length scales (b) if the Compton scale preserves its usual form. (c) and (d) are the corresponding diagrams if the duality between the Compton and Schwarzschild expressions is preservd.  From Ref.~\cite{lake}.
\label{MR2}
}
\end{figure}

Equation~(\ref{V1}) implies 
that the usual expression for the Schwarzschild radius no longer applies for masses below $M_E'$. If the black hole is assumed to be
spherically symmetric in the higher-dimensional space,  one has
\cite{kanti2004}
\begin{equation} \label{higherBH}
R_{\rm S} \simeq R_E  \left(\frac{M}{M_E'} \right)^{1/(n+1)} \, .
 \end{equation}
Therefore the slope of the black hole boundary in Fig.~\ref{MR} becomes shallower for $M < M_E'$,
as indicated in Fig.~\ref{MR2}(a).
The intersect with the
Compton line
then becomes
\begin{equation} \label{revisedplanck}
R_{\rm P}' \simeq (R_{\rm P}^2R_E^n)^{1/(2+n)}, \quad   
M_{\rm P}' \simeq (M_{\rm P}^2M_E^n)^{1/(2+n)} \,,
 \end{equation}
so $M_{\rm P}'  \ll M_{\rm P}$  and $R_{\rm P}'  \gg R_{\rm P}$ for $R_E  \gg R_{\rm P}$.

In principle, the lowering of the Planck mass could permit the possibility of TeV quantum gravity and the production of small black holes at the Large Hadron Collider (LHC), with their evaporation leaving a distinctive signature \cite{dimo}.
If the accessible energy is $E_{\rm max} \approx 10$~TeV, then the  extra dimensions can be probed for
\begin{equation} \label{nconstraint}
R_E  > 10^{-18 + 30/n}\, \mathrm{cm}
\simeq
\left\lbrace
\begin{array}{ll}
10^{12}\, \mathrm{cm} & (n=1 ) \\
10^{-3} \, \mathrm{cm} & (n=2) \\
10^{-14}\, \mathrm{cm} & (n=7) \, .
\end{array}
\right.
\end{equation}
Clearly, $n=1$ is excluded on empirical grounds but $n=2$ is possible. One expects $n=7$ in M-theory,
so it is interesting that $R_E$ must be of order a Fermi in this case.
One could also consider a scenario with a hierarchy of compactification scales, $R_i = \alpha_i \, R_{\rm P}$ with $\alpha_1 \geq \alpha_2 \geq .... \geq \alpha_n \geq 1$, such that the dimensionality progressively increases as one goes to smaller scales \cite{Ca:2013}. 
This situation is represented in Fig.~\ref{MR2}(b).  There is still no evidence for the extra dimensions \cite{atlas}, which suggests that either  they do not exist or they have a compactification scale $R_E$ which is so small that $M_{\rm P}'$ exceeds the energy attainable by the LHC.  

Another possible reason for the non-detection of accelerator black holes is that 
the $M$ dependence of $R_{\rm C}$ is also affected by the extra dimensions. 
Lake and myself have argued that the {\it effective}
 Compton wavelength 
depends on the form of the $(3+n)$-dimensional wavefunction \cite{lake}. 
If  this  is spherically symmetric in all the dimensions, then one  has $R_{\rm C} \propto M^{-1}$ (as usually assumed). However, if the wave function is pancaked in the extra dimensions and maximally asymmetric, then then we find $R_{\rm C} \propto M^{-1/(1+n)}$. 
This implies
that the duality between the Compton wavelength and the Schwarzschild radius 
persists in the higher dimensional case but
that there is no accelerator production of black holes. Thus the constraint on $R_E$ given by Eq.~(\ref{nconstraint}) no longer applies. 
This scenario is illustrated in Fig.~\ref{MR2}(c) for extra dimensions compactified on a single length scale $R_E$ and in Fig.~\ref{MR2}(d) for a hierarchy of length scales, when the extra dimensions help to smooth the minimum.
The latter case resembles the smooth minimum in Fig.~\ref{MR}, which suggests that higher dimensions might themselves underlie the BHUP correspondence.

The above discussion of higher-dimensional black holes has assumed that the simple power-law forms for $R_{\rm S}$ and $R_{\rm C}$ apply all the way to their intersect at the (modified) Planck scale. However, the BHUP correspondence  suggests that they should be unified in some way, which would smooth the minima in Figures~\ref{MR2}.  This raises the issue of the form of the GUP and BHUP correspondence in the higher-dimensional case.  If the Compton wavelength preserves its 3-dimensional form, one might expect the generalized  Compton wavelength to become 
 \begin{equation}
R _{\rm C}' =  \frac{\hbar }{Mc} \left[ 1  + \left( \frac{M}{M_{\rm P}'} \right)^{(n+2)/(n+1)} \right]  \quad (R < R_E) \, ,
\label{GUP5}
\end{equation}
so that $R _{\rm C}'$ becomes $R _{\rm S}'$ at large $M$.   If duality between $R_{\rm S}$ and $R_{\rm C}$ is preserved in the higher-dimensional case, one might expect 
\begin{equation}
R _{\rm C}' =  R_* \left(\frac{M_{\rm P}}{M }\right)^{1/(1+n)} \left[ 1  + \left( \frac{M}{M_{\rm P}'} \right)^{2/(n+1)} \right]  \quad (R < R_E) 
\label{GUP5}
\end{equation}
However, the literature on this gives different results \cite{kkmn}.

Finally, if we interpret the Compton wavelength as marking the boundary in the $(M,R$) diagram below which pair-production rates becomes significant, we might expect the presence of compact extra dimensions to affect pair-production rates at high energies. Specifically,  pair-production above the 
energy scale 
$M_E c^2 \equiv \hbar c/R_E$, should be enhanced relative to the 3-dimensional case. 
Indeed, there is tentative evidence that this is a generic feature of higher-dimensional theories 
~\cite{He99,Ebetal00}.

\section{Linking black holes and elementary particles}

The suggestion that there could be a fundamental link between elementary particles and black holes goes back to the 1970s, when it was motivated in the context of strong gravity theories \cite{sivaram}.  Various arguments supported this suggestion: (1) both hadrons and Kerr-Newman black holes are characterised by three parameters (M,J,Q);
(2) both have a magnetic dipole moment and a gyromagnetic ratio of 2 but no electric dipole moment; (3) Regge trajectories and extreme Kerr solutions have the same relationship between angular momentum and mass ($J \sim M^2$);  (4) when classical black holes interact, their surface area can never decrease, which is analogous to the increase in cross-sections found in hadron collisions.  

Of course,  elementary particles cannot be black holes with normal gravity, since their Compton wavelength is much larger than their Schwarschild radius, as illustrated in Fig.~\ref{MR}.  The early models therefore assumed that gravity becomes stronger by a factor of $G_F/G \sim (M_{\rm P}/m_p)^2 \sim 10^{38}$ on the hadronic scale.  This requires the existence of a massive spin-2 meson and corresponds to a short-range force.  If the hadronic resonances are extremal black holes, their mass and spin should satisfy $G_F m_h^2 = J$, corresponding to a Regge slope of $(1 {\rm Gev})^{-2}$, and they should have a spectrum of masses $M_n \sim n^{1/2}$ GeV \cite{oldershaw}. 

The current proposal -- explored in more detail in work with Mureika and Nicoloini \cite{CMN2022} -- is prompted by the Generalized Uncertainty Principle and the
duality between the Compton and Schwarzschild expressions under the transformation $M \rightarrow \Mpl^2/M$,  so the context is somewhat different. Also elementary particles are regarded as sub-Planckian black holes under normal gravity rather 
conventional black holes under strong gravity  However, 
there is a link with strong gravity because the force between two masses, while still obeying the inverse-square law,  is much enhanced for $M < \sqrt{\beta} \Mpl$.  

Extending the BHUP correspondence to charged 
black holes adds important insights.  Although
the Reissner-Nordstrom
itself has a nearly Planckian mass and therefore cannot represent an elementary particle, 
adding a GUP term introduces sub-Planckian solutions.  This explains why the charge cannot exceed $\sqrt{2 \beta/\alpha_e} \approx 1$ for $\beta \sim 10^{-2}$, as observed for elementary particles.  Similar considerations apply for spinning black holes. However, these solutions only correspond to extremal black holes 
if the mass is $\sqrt{\beta/2} \, M_P$, which  is too large for an elementary particle (given the allowed range of $\beta$).   

These considerations must be modified if there are extra dimensions on small scales.  Although there is some uncertainty in the modifications to the GUP in this case, 
Fig.~\ref{MR2} shows that the extra dimensions themselves smooth the Compton-Schwarzschild transition. 
Furthermore,  the black hole mass may be shifted down towards the hadron scale, the effective strength of gravity being increased by the extra dimensions.  However, the higher-dimensional analysis has not yet been extended to the charged and rotating black holes. 
Extra dimensions may also play an important role in amalgamating general relativity and quantum theory,  with higher-dimensional relativity permitting a classical-type interpretation of some quantum anomalies. 

\begin{acknowledgement}
I thank my collaborators in the work reported here: Matthew Lake, Heather Mentzner, Leonardo Modesto, Jonas Mureika, Piero Nicolini and Isabeau Pr\'emont-Schwarz.  
\end{acknowledgement}

\end{document}